\documentclass{iau_FM}
\usepackage{graphicx}

\title[S320.~~6 years of Heliometer in RJ] 
{The Reflecting Heliometer of Rio de Janeiro after 6 Years of Activity}

\author[Boscardin, Sigismondi, Penna, D'Avila, Reis-Neto \& Andrei] 
{S. C. Boscardin$^1$, C. Sigismondi$^2$, J. L. Penna$^1$, V. D'Avila$^3$, E. Reis-Neto$^4$ \and A.H. Andrei$^{1,5,6}$}  

\affiliation{$^1$ON-MCTI, RJ/BR, $^2$ICRA/Sapienza, Roma/IT, $^3$ UERJ, RJ/BR, $^4$MAST-MCTI, RJ/BR, $^5$ OV-UFRJ, RJ/BR, $^6$ SYRTE/OP, Paris/FR \\ 
email: {\tt sergio.boscardin@on.br , sigismondi@icra.it}}

\pubyear{2015}
\volume{S320}
\jname{Solar and Stellar Flares and Their Effects on Planets} 
\editors{A. Kosovichev, S. Hawley \& P. Heinzel, eds.}
\begin{document}

\maketitle

\begin{abstract}
Started its regular, daily operational phase in 2011 and installed in 2009 by the occasion of the Symp264 in the XXVII IAU GA at Rio de Janeiro, the results so far obtained show that the Heliometer of the Observat\'orio Nacional fulfilled its planed performance of single measurement to the level of few tens of milli-arcsecond, freely pivoting around the heliolatitudes without systematic deviations or error enhancement.
We present and discuss the astrometric additions required on ground based astronomic programs. We also discuss instrumental and observations terms, namely the constancy of the basic heliometric angle, against which the measurements are made, and the independence to meteorological and pointing conditions. 
\keywords{Solar Diameter, methods: data analysis, instrumentation: high angular resolution, Sun: fundamental parameters, Sun: general.}
\end{abstract}

\firstsection 
\section{Introduction: from the Solar Astrolabes to the Reflecting Heliometer}

The Reflecting Heliometer was conceived at the Observatorio Nacional to measure the solar diameter with metrological accuracy and to follow
the long observational campaigns started since 1974 in S\~ao Paulo University and Nice/Calern Observatory with the Astrolabes of Danjon, adapted to solar observations (\cite[Andrei et al. (2015)]{Andrei15}).
While the astrolabes could measure by timing the drift-scan of the vertical diameter, to which at each time of the day and season corresponded a given heliolatitude, the Heliometer can measure the solar diameter at all heliolatitudes by rotating around its axis.
Moreover each drift-scan of the solar disk across an almucantarat of given altitude ranged from 2 to 6 minutes, while a measurement of the Heliometer is the result of an average of 50 images taken in 10 seconds, without the limitation of the meridian transit, off limits for the astrolabes while the solar figure is less affected by refraction and turbulence.   
The Reflecting Heliometer eliminated the dependence on wavelengths of the refracting surfaces in the classical instrument of Fraunhofer and in the version of Goettingen with a front prism made by Schur and Ambronn in 1890-5, \cite[Meyermann (1939)]{Meyermann39}.
The number of observations along the regular sessions started in 2011 increased of a factor of 20 with respect to the Astrolabe.
The apparatus, and the results of the regular daytime observations, amounting to thousands of measurements per year are reviewed in \cite[Andrei et al. (2014)]{Andrei14}. High order astrometric terms in the data reduction, namely (a) The second order terms for diurnal aberration and parallax; these effects are found negligible on themselves, and also on the building up of spurious yearly trends. On the contrary, for the accounting of (b)the Earth´s ellipticity of the orbit the standard astrometric procedures had to be upgraded to make room for the full description of daily variations, to the mas level, instead of the usual approximation to the mean observational day. A thorough model for (c) the second order atmospheric refraction has had to be developed to match the systematics left on the observations by using the geometrical second order description. 

\section{Measures accuracy, stability and calibration of the heliometric angle}
The large number of measurements available with the reflecting heliometer at all heliolatitudes allows to reduce the errorbars by statistics. The accuracy $\sigma \propto 1/\sqrt(N)$ where $\sigma\sim 0.5''$ is the accuracy obtained from the single set of 50 images at ordinary atmospheric seeing of $2 \sim 3 ''$ and $\ge 100 $Hz frequency averaged over 100 cycles.
The fluctuations around the true value of the solar diameter, due to Earth's atmosphere motions, are considered as random and the accuracy of 100 mas is obtained usually in a single observative session of 1 hour under clear sky.

Measures variations of  $\sigma =1\div 2$ arcsec, thus larger than the statistical errors, occurred three times after the glass filter replacement.
These variations were found as dependant on the position of the case of the filter, leading to identify the prismatic effect of the filter duplicating the image of the Sun. The second image was super-imposed to the first one with a shift of a fraction of arcsecond, leading to a falsed measurement of the gap between the edges of the images produced by the splitted mirrors: the gap is reduced of the same shift, and this explain the abrupt variations of the measured diameter after moving the filters, \cite[Sigismondi et al. (2015)]{Sigismondi15}. 
Afterwards a mylar filter ensured the elimination of prismatic effects to the images.
The stability of the heliometric angle has been checked along the years of operation with the glass filters never moved for two years and, after, with mylar filters always fixed because of their case. This angle has been measured also with a fixed wooden reference rod, with special pinhole masks to focus the rod at finite distance on the focal plane of the telescope designed for sources at infinity. 
The Reflecting Heliometer was carried to Easter Island for the total eclipse of 2009 to calibrate the heliometric angle with the Baily beads method \cite[Sigismondi (2009)]{Sigismondi09}. After the least squares method applied to the measures made in periods with the same filter's configuration have been compared with corresponding eclipse data showing the stability of the configuration. Changing the filter's positions the effective heliometric angle changed, it is the combination of the mirrors' angle, the filter's prismatic effect and a mechanical stress of the telescope at the insertion of the filter's case. A relaxation time was experienced before recovering the diameter's measurements with the new effective heliometric angle. 
The systematic shift of 1.3 arcsec from the treated series (average value 958.7 arcsec, fig. 1) to the actual solar diameter in visual band of 960.0 arcsec is recently confirmed by the determination of the solar diameter with total eclipses, \cite[Lamy et al. (2015)]{Lamy15}, and the 2012 transit of Venus, \cite[Sigismondi et al. (2015b)]{Sigismondi15b}.

\begin{figure}[!tbp]
\includegraphics[width=5.2in]{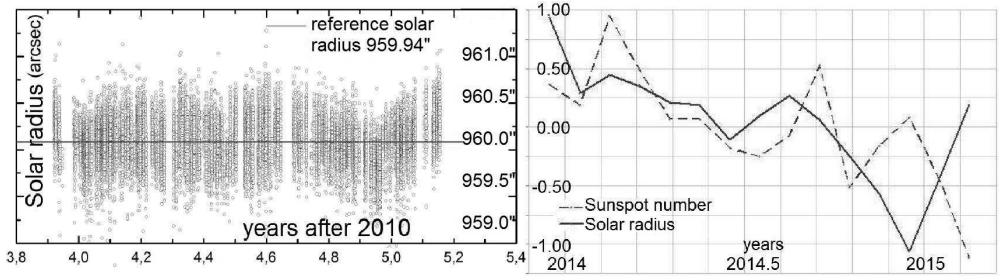}
\caption{On the left, the Heliometer series of measured photospheric semi-diameter corrected for 1) linear trend in time 2) solar longitude (annual term) 3) tangent of zenital distance (refraction) 4) angle of the mirrors' splitting line (rotation of the telescope's axis to measure different heliolatitudes) 5) adjustment of the average value to the actual reference solar radius of 959.94" (horizontal line). On the right, the running monthly averages of the solar radius (continuos line) are compared to the equivalent monthly averages of the sunspots count (dotted lines), with evident agreement.}
\end{figure}

%

\section{The Heliometric and the Solar activity series}

After correcting for the instrumental, observational, and astrometry terms demanded for the mas-level final accuracy, the series of semi-diameter monitoring becomes clearly much more stable, as shown in fig1a. At the same time, the series is clearly off from constant. As done for earlier measurements, we collate the resulting series with the magnetic solar activity as given by the daily Sunspot Number. To compare the variations of solar semi-diameter with the magnetic solar activity we normalized both the series, in other words we diminished all the values from the average and divided the result by the standard deviation. With the obtained numbers we made monthly averages. We verify that the two series show a strong correlation witch can be seen in fig1b.

\section{Conclusions}

The Reflecting Heliometer at the Observat\'orio Nacional in Rio de Janeiro is routinely performing thousands of observations of the solar diameter with an accuracy of few tenths milliarcseconds, on daily basis. This result performed in all heliolatitudes, sets a new groundbased standard in solar metrology. The series of data 2010-2015 has been treated for the major astrometric corrections and can be used to infer the behavior of the solar diameter at the end of cycle 23 beginning 24, after the deepest solar minimum of the last two centuries.
The heliometric angle has been verified stable along the years to the level required for solar metrology, and the role of the glass filter has been studied and its prismatic effect, responsible of the gaps observed in the solar radius after changing the positions of the filter, has been eliminated.  
The heliometric measurements of the solar diameter are comparable in precision with the satellite ones, and are possible under clear skies at Rio Observatory. In 2015 in the first 142 days there were 85 duty days with 28 cloudy and 1 day of maintenance: 4479 measurements have been done, with an average of 75 measurements per observing day and more than 11500 per calendar year. The previous observations made with the Astrolabe in the period 1998-2009 were 21000 in total. Usually these measurements are carried by the same astronomer S. C. Boscardin, but the procedure of data acquisition and analysis is totally unpersonal after the pointing and the consensus to observe.
The possibility to study the variations of the solar diameter in visual band with a time resolution of one day and space angular resolution of 10 mas is the highlight of the Reflecting Heliometer, it allows to monitor in real time the solar activity in diameter's pulses to be compared with the spots or Coronal Mass Ejections activity with a continuity and homogeneity not possible with space instruments.

\end{document}